# PEGylated Nano-Graphene Oxide for Delivery of Water Insoluble Cancer Drugs

Zhuang Liu, Joshua T. Robinson, Xiaoming Sun and Hongjie Dai*

*Department of Chemistry, Stanford University, Stanford, CA, 94305, USA*

Email: hdai@stanford.edu

Graphene has emerged as a 2D material with interesting physical properties.[1,2] Intensive research is on-going to investigate the quantum physics in this system and potential applications for nano-electronic devices[2], transparent conductors and nano-composite materials[3]. Thus far, little has been done to explore graphene in biological systems, despite much effort in the area of carbon nanotubes for in vitro and in vivo biological applications.[4-9] Here, we synthesize and functionalize nanoscale graphene oxide (NGO) sheets (<50nm) by branched, biocompatible polyethylene glycol (PEG) to render high aqueous solubility and stability in physiological solutions including serum. We then uncover a unique ability of graphene in attaching and delivery of aromatic, water insoluble drugs.

It is known that clinical use of various potent, hydrophobic molecules (many of them aromatic) is often hampered by their poor water solubility. Although synthesis of water soluble pro-drugs may circumvent the problem, the efficacy of the drug decreases. Here, we show that PEGylated NGO (NGO-PEG) readily complexes with a water insoluble aromatic molecule SN38, a camptothecin (CPT) analog,[10] via non-covalent van der Waals interaction. The NGO-PEG-SN38 complex exhibits excellent aqueous solubility and retains the high potency of free SN38 dissolved in organic solvents. The toxicity exceeds that of irinotecan (CPT-11, a FDA approved SN38 prodrug for colon cancer treatment) by 2-3 orders of magnitude.

We prepared graphene oxide by oxidizing graphite using a modified Hummer's method.[3,11] The resulting GO (single layered and few-layered, Supp Info. Fig.S1) was soluble in water but aggregated in solutions rich in salts or proteins such as cell medium and serum (Fig. 1a). This was likely due to screening of the electrostatic charges and non-specific binding of proteins on the GO.[12] To impart aqueous stability and prevent bio-fouling, we sonicated the GO to make them into small pieces and conjugated a 6-armed PEG-amine stars to the carboxylic acid groups[3] (Supp Info Fig. S3) on GO via carbodiimide catalyzed amide formation. The resulting PEGylated NGO exhibited excellent stability in all biological solutions tested including serum (Fig. 1b). PEGylation was further confirmed by infrared (IR) spectroscopy (Supp Info. Fig.S1b). The as-made GO sheets were 50-500nm in size (Fig.1c), whereas NGO-PEG was ~5-50 nm (Fig. 1d) due to sonication steps (see Supp Info.).

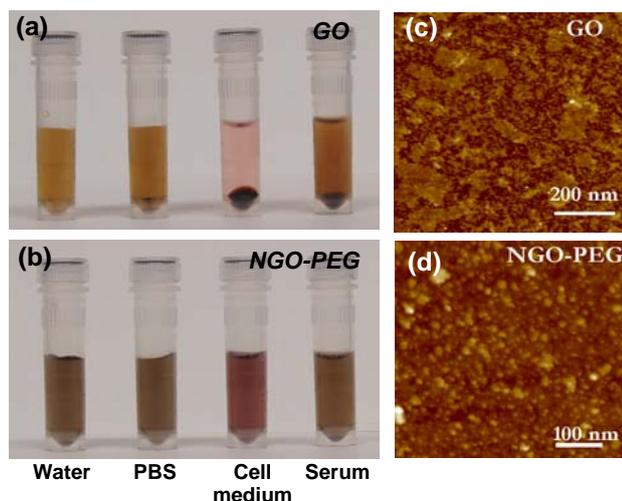

*Figure 1.* PEGylation of graphene oxide. (a&b), photos of GO (a) and NGO-PEG (b) in different solutions recorded after centrifugation at 10,000 g for 5 minutes. GO crashed out slightly in PBS and completely in cell medium and serum (top panel). NGO-PEG was stable in all solutions. (c&d) AFM images of GO (c) and NGO-PEG (d).

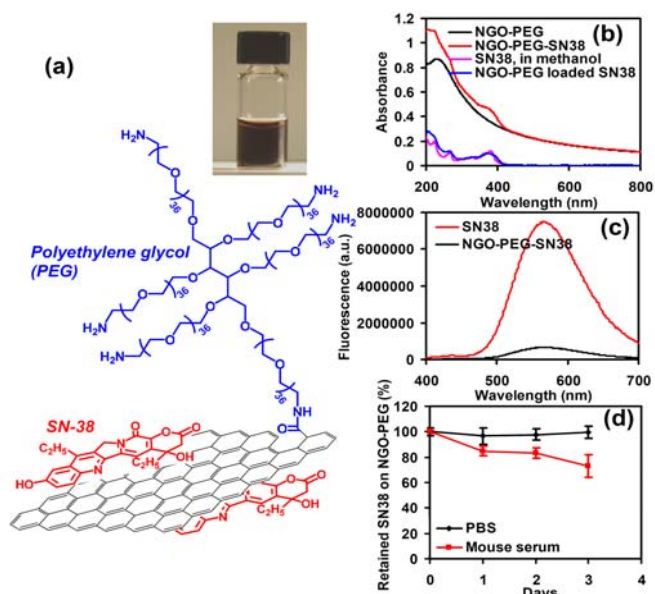

*Figure 2.* SN38 loading on NGO-PEG. (a) schematic draw of SN38 loaded NGO-PEG. Inset: a photo of NGO-PEG-SN38 water solution. (b) UV-VIS absorption spectra of NGO-PEG, NGO-PEG-SN38, SN38 in methanol and difference spectrum of NGO-PEG and NGO-PEG-SN38. The SN38 absorbance at 380 nm was used to determine the loading. (c) Fluorescence spectra of SN38 and NGO-PEG-SN38 at [SN38]=1μM. Significant fluorescence quenching was observed for SN38 adsorbed on NGO. (d) Retained SN38 on NGO-PEG over time incubated in PBS and serum respectively. SN38 loaded on NGO-PEG was stable in PBS and released slowly in serum. Error bars were based on triplet samples.

We then investigated the binding of SN38 to NGO-PEG. We chose SN38 as a cargo because SN38 is a potent topoisomerase I inhibitor.[10] To be active, CPT-11 currently used in clinic has to be metabolized to SN38 after systematic adminsitration.[10,13] However a large amount of CPT-11 is excreted before transforming to SN38 or metabolized to other inactive compounds.[14] The water insolubility has prevented the direct use of SN38 in the clinic.[10]

We found that SN38 was complexed with NGO-PEG (Fig.2a) by simple mixing of SN38 dissolved in DMSO with a NGO-PEG water solution. The excess, uncoupled SN38 precipitated and was removed by centrifugation. Repeated washing and filtration were used to remove DMSO and any residual free SN38 (see Supp Info. for details). UV-VIS spectrum of the resulting solution revealed SN38 peaks superimposing with the absorption curve of NGO-PEG (Fig. 2b), suggesting loading of SN38 onto NGO-PEG. Based on the extinction coefficients, we estimated that 1 gram of NGO-PEG loaded ~0.1 gram of SN38 (Supp Info.). An increase in sheet thickness was observed after SN38 loading on NGO-PEG (Supp Info. Fig. S3). A control experiment revealed no loading of SN38 on PEG polymer in a solution free of NGO.

Unlike free SN38, which was very insoluble in water, NGO-PEG-SN38 complexes were water soluble at concentrations up to ~1 mg/mL (in terms of SN38). Fluorescence spectra of NGO-PEG-SN38 and free SN38 at the same SN38 concentration showed drastic fluorescence quenching of SN38 in the NGO-PEG-SN38 case (Fig.2c), suggesting close proximity of SN38 to the NGO sheets. We suggest that binding of SN38 onto NGO-PEG was non-covalent in nature, driven by hydrophobic interactions and π- π stacking[15,16] between SN38 and aromatic regions of the graphene oxide sheets. The existence of aromatic conjugated domains on GO was revealed by NMR previously[17]

To determine the stability of SN38 loaded on NGO-PEG and release rate, we incubated NGO-PEG in phosphate buffer saline (PBS) and mouse serum respectively at 37 ºC and measured the percentage of retained SN38 on NGO-PEG (Fig. 2d, See Supp Info. for experimental details). We found that SN38 on NGO-PEG exhibited negligible release from NGO in PBS and ~30% release in serum in 3 days (Fig.2d). This suggested strong non-covalent binding of SN38 on graphene oxide sheets. The slow but finite release of SN38 in serum was likely caused by the binding of SN38 by serum proteins,[18] useful for drug delivery.

MTS assay found that NGO-PEG-SN38 afforded highly potent cancer cell killing in vitro with a human colon cancer cell line HCT-116. The water soluble drug CPT-11 was found to be the least toxic, with a 50% growth inhibition concentration (IC50) of ~10 µM (Fig. 3). Our water soluble NGO-PEG-SN38 exhibited high potency with IC50 values of ~6nM for HCT-116 cells, which is ~ 1000 fold more potent than CPT-11 and similar to that of free SN38 dissolved in DMSO (Fig. 3a). The high potency of NGO-PEG-SN38 was also observed with various other cancer cell lines tested (Supp Info. Table S1). Importantly, no obvious toxicity was measured for various concentrations of plain NGO-PEG without drug loading (Fig.3b), suggesting that the PEGylated nanographene oxide sheets were not cytotoxic by themselves. Apoptosis assay further confirmed no obvious increase of cell death or apoptosis after incubating cells with plain NGO-PEG (Supp Info. Fig. S 4). The cellular uptake of NGO-PEG was likely via endocytosis as evidenced by confocal fluorescence microscopy data (Supp Info. Fig. S5).

We found that the strategy of attaching various types of insoluble, aromatic drug molecules onto NGO-PEG via simple adsorption was general. Other drugs that we succeeded in loading onto NGO-PEG by simple adsorption included different camptothecin analogs and Iressa (geftinib), an potent epidermal growth factor receptor (EGFR) inhibitor (Supp Info. Fig. S6&7).

The water soluble NGO-PEG-SN38 complex could open up a window to potential use of this drug. Graphitic nanocarriers including nanographene sheets and carbon nanotubes afford strong noncovalent binding with aromatic drugs via simple adsorption.[16] Graphene sheets as drug carrier are interesting because both sides of a single sheet could be accessible for drug binding. The unique 2D shape and ultra-small size (down to 5 nm) of NGO-PEG may offer interesting in vitro and in vivo behaviors. Moreover, the low cost and large production scale of graphite and GO is unmatched by carbon nanotubes. Thus the biocompatible nanographene sheets are novel materials promising for biological applications.

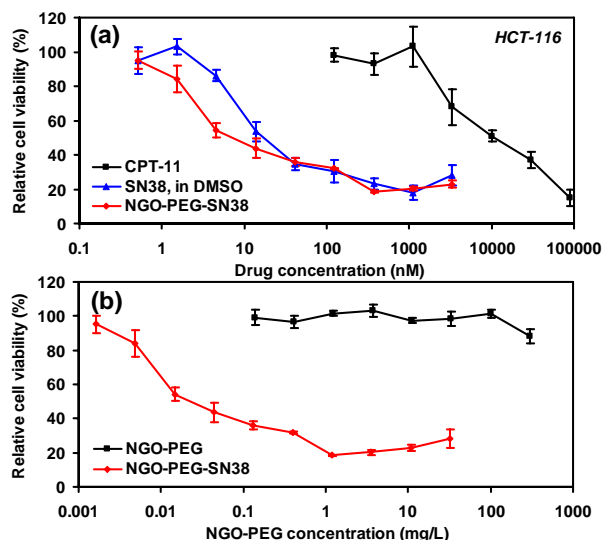

*Figure 3*. In vitro cell toxicity assay. (a) Relative cell viability (versus untreated control) data of HCT-116 cells incubated with CPT-11, SN38 and NGO-PEG-SN38 at different concentrations for 72 h. Free SN38 was dissolved in DMSO and diluted in PBS. Water soluble NGO-PEG-SN38 showed similar toxicity as SN38 in DMSO and far higher potency than CPT-11. (b) Relative cell viability data of HCT-116 cells after incubation with NGO-PEG with (red) and without (black) SN38 loading. Plain NGO-PEG exhibited no obvious toxicity even at very high concentrations. Error bars were based on triplet samples.

**Acknowledgement:** This work was partly supported by NIH-NCI CCNE-TR, a Stanford BioX Grant and a Stanford graduate fellowship.

**References**
(1) Geim, A. K.; Novoselov, K. S. *Nat. Mater.* **2007**, *6*, 183-191.
(2) Li, X. L.; Wang, X. R.; Zhang, L.; Lee, S. W.; Dai, H. J. *Science* **2008**, *319*, 1229-1232.
(3) Stankovich, S.; Dikin, D. A.; Dommett, G. H. B.; Kohlhaas, K. M.; Zimney, E. J.; Stach, E. A.; Piner, R. D.; Nguyen, S. T.; Ruoff, R. S. *Nature* **2006**, *442*, 282-286.
(4) Kam, N. W. S.; Jessop, T. C.; Wender, P. A.; Dai, H. J. *J. Am. Chem. Soc.* **2004**, *126*, 6850-6851.
(5) Kam, N. W. S.; O'Connell, M.; Wisdom, J. A.; Dai, H. *Proc. Natl. Acad. Sci. USA* **2005**, *102*, 11600-11605.
(6) Liu, Z.; Cai, W. B.; He, L. N.; Nakayama, N.; Chen, K.; Sun, X. M.; Chen, X. Y.; Dai, H. J. *Nat. Nanotech.* **2007**, *2*, 47-52.
(7) Liu, Z.; Davis, C.; Cai, W.; He, L.; Chen, X.; Dai, H. *Proc. Natl. Acad. Sci. USA* **2008**, *105*, 1410-1415.
(8) Cherukuri, P.; Gannon, C. J.; Leeuw, T. K.; Schmidt, H. K.; Smalley, R. E.; Curley, S. A.; Weisman, R. B. *Proc. Natl. Acad. Sci. USA* **2006**, *103*, 18882-18886.
(9) Feazell, R. P.; Nakayama-Ratchford, N.; Dai, H.; Lippard, S. J. *J. Am. Chem. Soc.* **2007**, *129*, 8438-8439.
(10) Tanizawa, A.; Fujimori, A.; Fujimori, Y.; Pommier, Y. *J. Natl. Cancer Inst.* **1994**, *86*, 836-842.
(11) Hummers, W. S.; Offeman, R. E. *J. Am. Chem. Soc.* **1958**, *80*, 1339-1339.
(12) Kam, N. W. S.; Dai, H. J. *J. Am. Chem. Soc.* **2005**, *127*, 6021-6026.
(13) Kawato, Y.; Aonuma, M.; Hirota, Y.; Kuga, H.; Sato, K. *Cancer Res.* **1991**, *51*, 4187-4191.
(14) Slatter, J. G.; Schaaf, L. J.; Sams, J. P.; Feenstra, K. L.; Johnson, M. G.; Bombardt, P. A.; Cathcart, K. S.; Verburg, M. T.; Pearson, L. K.; Compton, L. D.; Miller, L. L.; Baker, D. S.; Pesheck, C. V.; Lord, R. S. *Drug Metabolism and Disposition* **2000**, *28*, 423-433.
(15) Chen, R.; Zhang, Y.; Wang, D.; Dai, H. J. *J. Am. Chem. Soc.* **2001**, *123*.
(16) Liu, Z.; Sun, X.; Nakayama, N.; Dai, H. *ACS Nano* **2007**, *1*, 50-56.
(17) Lerf, A.; He, H. Y.; Forster, M.; Klinowski, J. *J.Phys. Chem. B* **1998**, *102*, 4477-4482.
(18) Burke, T. G.; Mi, Z. H. *J. Med. Chem.* **1994**, *37*, 40-46.